\documentclass[10pt,twocolumn]{article}

\usepackage[T1]{fontenc}
\usepackage[utf8]{inputenc}
\usepackage{mathptmx}      
\usepackage[scaled=0.92]{helvet}  
\usepackage[english]{babel}
\usepackage[letterpaper,margin=0.85in,columnsep=0.28in]{geometry}
\usepackage{graphicx}
\graphicspath{{figures/}{./}}
\usepackage{xcolor}
\definecolor{accent}{gray}{0}
\definecolor{accentdark}{gray}{0}
\definecolor{sagerule}{gray}{0}
\definecolor{rulegrey}{gray}{0.55}
\definecolor{addedblue}{RGB}{0,0,205}

\usepackage{pgfplots}
\pgfplotsset{compat=1.18}
\usepgfplotslibrary{fillbetween}
\definecolor{steelblue}{RGB}{70,129,180}
\definecolor{lightfill}{RGB}{197,227,237}
\definecolor{quietshade}{gray}{0.82}
\definecolor{warmgrey}{RGB}{110,108,102}
\definecolor{darkbluegrey}{RGB}{45,75,95}
\usepackage{amsmath}
\usepackage{amssymb}
\usepackage{booktabs}
\usepackage{array}
\usepackage{caption}
\usepackage{ragged2e}
\usepackage{microtype}
\usepackage{enumitem}
\usepackage{titlesec}
\usepackage{fancyhdr}
\usepackage[hidelinks]{hyperref}
\hypersetup{colorlinks=false,allbordercolors=white}
\usepackage[round]{natbib}

\microtypesetup{protrusion=true,expansion=true}

\titleformat{\section}
  {\normalfont\large\bfseries\sffamily\color{accent}}
  {\color{accent}\thesection}{0.6em}{}
\titleformat{\subsection}
  {\normalfont\normalsize\bfseries\sffamily\color{accentdark}}
  {\color{accentdark}\thesubsection}{0.6em}{}
\titleformat{\subsubsection}
  {\normalfont\normalsize\itshape\sffamily\color{accentdark}}
  {\color{accentdark}\thesubsubsection}{0.6em}{}
\titlespacing*{\section}{0pt}{1.5ex plus 1ex minus .2ex}{0.9ex plus .2ex}
\titlespacing*{\subsection}{0pt}{1.2ex plus 1ex minus .2ex}{0.6ex plus .2ex}
\titlespacing*{\subsubsection}{0pt}{1.0ex plus 1ex minus .2ex}{0.5ex plus .2ex}

\usepackage{lastpage}
\renewcommand{\headrulewidth}{0pt}
\renewcommand{\footrulewidth}{0.4pt}
\renewcommand{\footrule}{\hbox to\headwidth{\color{rulegrey}\leaders\hrule height \footrulewidth\hfill}}
\renewcommand{\headrulewidth}{0.4pt}
\renewcommand{\headrule}{\hbox to\headwidth{\color{rulegrey}\leaders\hrule height \headrulewidth\hfill}}
\fancypagestyle{plain}{%
  \fancyhf{}%
  \renewcommand{\headrulewidth}{0pt}%
  \renewcommand{\footrulewidth}{0pt}%
  \fancyfoot[C]{\footnotesize\sffamily\color{rulegrey}\thepage}%
}
\makeatletter
\newcommand{\sagetitle}[1]{\gdef\@sagetitle{#1}}
\makeatother

\newcommand{\SAGEfront}{%
  \begin{center}
    {\sffamily\bfseries\fontsize{16}{19}\selectfont\color{accentdark}
     Credible, Not Always Correct:\\[2pt]
     How Reddit Users Verify AI-Generated Legal Advice\par}
    \vspace{1.0em}
    {\fontsize{11}{13}\selectfont
     Rebecca~Owens\textsuperscript{\textcolor{accent}{1,\,$\ast$}},
     Yusuf~Mücahit~Çetinkaya\textsuperscript{\textcolor{accent}{2}, \textcolor{accent}{3}},
     Stergios~Aidinlis\textsuperscript{\textcolor{accent}{1}},
     Dhyey~Mehta\textsuperscript{\textcolor{accent}{2}},
     and Tuğrulcan~Elmas\textsuperscript{\textcolor{accent}{2}}\par}
    \vspace{0.5em}
    {\footnotesize
     \textsuperscript{1}Durham Law School, Durham University, UK\quad
     \textsuperscript{2}School of Informatics, University of Edinburgh, UK \par
     \textsuperscript{3}Department of Computer Engineering, Middle East Technical University, Türkiye
     \par}
  \end{center}
  \vspace{0.3em}
}

\begin{document}
\thispagestyle{plain}

\twocolumn[
  \begin{@twocolumnfalse}
  \SAGEfront
  \begingroup
  \setlength{\leftskip}{0.055\textwidth}
  \setlength{\rightskip}{0.055\textwidth}
  {\noindent\sffamily\bfseries\small Abstract\par}
  \vspace{0.35em}
  {\small
  Large language models (LLMs) are increasingly used by laypeople to resolve real legal problems, against a backdrop of persistent access-to-justice deficits. This article presents evidence that the practical force of AI-generated legal advice depends not on its accuracy but on the social production of its credibility. While existing research has assessed the accuracy of legal AI, less is known about how machine-generated guidance is verified and made credible enough for lay users to act on. Drawing on a dual-method analysis of 153 Reddit narratives and 5{,}341 community reactions, this article maps a spectrum of verification practices. At one end, a minority of users verify AI-generated legal advice by triangulating across models, and some submit AI-generated guidance to platform communities for evaluation before acting, a configuration we term distributed counsel. Far more commonly, however, narratives are silent on verification. AI-generated legal advice is acted on the strength of its lawyer-like form and emotional reassurance alone. These findings show that AI-assisted legal self-help operates within an emerging informal infrastructure which redistributes the work of verification to those least equipped to bear it.\par}
  \vspace{0.7em}
  {\noindent\sffamily\bfseries\small Keywords\par}
  \vspace{0.25em}
  {\small Large Language Models (LLMs), Legal Technology, ChatGPT, Legal Advice, Access to Justice, Reddit, Digital Infrastructures\par}
  \endgroup
  \vspace{1.0em}
  \vspace{0.9em}
  {\footnotesize
   \noindent\textsuperscript{$\ast$}\,\textbf{Corresponding author:} Rebecca Owens, Durham Law School, Durham University, UK. Tel.: +44 191 334 2800. Email: \href{mailto:rebecca.owens@durham.ac.uk}{\texttt{rebecca.owens@durham.ac.uk}}\par}
  \vspace{1.2em}
  \end{@twocolumnfalse}
]

\section{Introduction}

Large language models (LLMs) such as ChatGPT are known to hallucinate citations, cloak errors in legalese, and operate outside the regulatory regimes that govern professional legal services \citep{cheong2024, munir2025}. Yet laypeople are turning to these systems for real legal advice in legal proceedings. Drawing on a computational social science analysis of Reddit accounts in which lay users describe using AI-generated legal advice for real-world issues, this article asks how this happens. It also aims to understand what this reveals about the social organization of legal credibility when professional counsel is just out of reach.

This article presents evidence suggesting that the practical force of AI-generated legal advice derives not from its accuracy but from the form of its outputs and the platform dynamics, and that this force operates, in most cases, without any verification at all. Framed this way, the problem we tackle in this article belongs to a long tradition of scholarship on how legality is produced and experienced in everyday life rather than in courtrooms \citep{sandefur2022}. Research shows that an individual's capacity to manage these encounters depends on their knowledge and resources \citep{mcdonald2021}. When access to a lawyer becomes unaffordable and difficult to access, the need for credible legal advice does not disappear. Instead, it is resolved elsewhere, through different actors and arrangements. Our central claim is that LLMs such as ChatGPT and online communities are becoming part of those arrangements. Therefore, understanding how AI-generated legal advice acquires practical force for lay legal users requires studying whether, how, and by whom that advice is verified, not merely auditing model accuracy.

Within the formal legal system, the credibility of advice is institutionally produced, and an accepted pre-existing information asymmetry exists between clients and lawyers \citep{chaserant2013}. Citizens need not assess the credibility of guidance themselves because licensure and ethical obligations underpin it \citep{simshaw2023}. If anything goes wrong, lawyers may have breached various ethical codes and are liable for damages. Yet what, if anything, performs this credentialing function when lay users turn to LLMs? Our evidence identifies two answers. In a minority of cases, platform communities step into the validation role, in a configuration we term \emph{distributed counsel}. This is where an LLM generates advice, a lay user applies it, and a community scrutinizes it, flagging errors and contesting overconfident guidance. Far more commonly, Reddit users act with no verification reported. AI-generated legal advice is relied upon because it reads as lawyer-like and is delivered in a reassuring manner. As we will show, this lack of verification is the result of a quiet relocation of legal authority to textual form, shifting the burden of verification of the advice onto users least equipped to bear it.

The conditions that push people toward these tools are well documented. Access to formal legal advice has traditionally been mediated by a tightly regulated professional framework that imposes ethical obligations and accountability to protect clients and ensure the integrity of services \citep{wald2022}. Yet the prohibitive cost of legal services has rendered this framework inaccessible to a significant proportion of lay users \citep{hadfield2014, sandefur2022}, producing what may be understood as an infrastructural deficit. This is a condition in which formally available legal systems cannot be accessed or used in practice by those who depend on them.

To characterize this deficit as infrastructural is to draw on an evolving body of scholarship showing that digital platforms can assume infrastructural functions \citep{plantin2018, cohen2024}. Following \citet{star1994}'s definition, infrastructure is understood not as a fixed set of arrangements but as a relational and embedded system that becomes visible at the point of breakdown. Where legal expertise is too costly or procedurally complex, the legal system ceases to function as usable infrastructure, and alternative systems are repurposed to perform infrastructural roles.

LLMs have emerged as one such alternative arrangement. Their appeal rests on a distinctive combination of low-cost or free access and perceived expert-like knowledge \citep{perlman2023, ryan2024, kant2024}. These features are often framed as democratizing access to justice, yet access to the most specialized legal systems remains uneven. Although increasingly sophisticated legal LLMs are emerging, they remain largely restricted to legal professionals \citep{lorek2024}. General-purpose systems such as ChatGPT and Gemini therefore constitute the de facto legal information environment for many lay users who cannot readily access formal legal services. For these users, accessibility and perceived expertise may generate substantial trust in AI-generated legal advice and, in some cases, lead them to privilege it over professional guidance \citep{schneiders2025}. Unlike traditional legal services, however, these systems provide no institutional basis for that trust. Users must therefore assess the credibility of AI-generated advice themselves, despite often lacking the legal expertise required, a difficulty compounded by hallucinations and the persuasive fluency of legalistic language \citep{munir2025}.

Existing research has largely framed the risks of AI-generated legal advice in terms of accuracy and automation bias \citep{magesh2025, nielsen2024, tan2023}. These approaches are necessary but incomplete. They explain whether an output is wrong, or why users may defer excessively to an automated system, but not how machine-generated language comes to acquire socially recognized legal credibility outside a professional advice relationship. Scant evidence largely derives from survey-based attitudes \citep{seabrooke2024} that are constrained by self-report bias. Such methods are ill-suited to capturing situated, real-time interactions with LLMs. Using Reddit data helps close this gap. Its pseudonymity may encourage users to disclose sensitive legal experiences, while its topic-specific subreddits and visible credibility signals render the evaluation of AI-generated guidance observable as a collective epistemic process \citep{shen2026}. In this article, we apply computational social science methods to analyze three research questions:

\begin{enumerate}[leftmargin=1.4em,itemsep=2pt,topsep=3pt]
  \item How commonly, and through what practices, do lay users report verifying AI-generated legal advice before acting on it?
  \item How do platform communities respond when AI-generated legal advice is shared, and how does that scrutiny vary across legal and technology-focused forums?
  \item In what legal domains, through what AI roles, and with what narrated outcomes do lay users act on AI-generated legal advice?
\end{enumerate}

Our findings reveal a structural mismatch at the heart of this practice. AI-generated legal advice arrives already bearing the marks of credibility. It is fluent, legally specific, professionally expressed, yet the burden of judging and validating the information falls on the party least equipped to carry it. In the regulated advice relationship, the actor who produces legal advice also bears responsibility for its soundness. When relying on AI-generated legal advice, that alignment breaks down. The LLM that generates the guidance may have no responsibility for the output \citep{pandit2026}. This leaves the lay user, who is not legally trained and cannot assess the credibility of the advice, to bear both the decision and the risk. The main question becomes not whether the model is right, but whether anyone ever checks.

This redistribution occurs within an emerging informal infrastructure of legal self-help. Digital platforms may assume infrastructural functions when they shape access, visibility, participation, and evaluation in areas historically mediated through public or professional institutions \citep{plantin2018, cohen2019}. In the context examined here, LLMs assist with drafting, legal research, document review, strategic planning, and emotional support, while Reddit communities provide a layer of scrutiny. The result is not a replacement for professional legal services, but a redistribution of legal work, credibility assessment, and risk across users and platforms.

The remainder of the paper proceeds as follows. Section~\ref{sec:background} develops the theoretical framework for analyzing AI-assisted legal advice and the production of legal credibility outside formal advice relationships. Section~\ref{sec:methods} sets out the computational social science methods and ethical safeguards adopted, while Section~\ref{sec:results} presents the empirical findings on how AI-generated legal advice is validated and used and how platform communities respond to it. Section~\ref{sec:discussion} draws these findings together to argue that LLMs are reshaping the conditions under which legal advice acquires credibility and practical force by decoupling apparent legal authority from institutional verification and redistributing responsibility for scrutiny to lay users and platform communities. Section~\ref{sec:conclusion} concludes by arguing that, if these patterns extend beyond Reddit users, the growing use of LLMs for legal self-help may widen access to legal assistance while simultaneously shifting the burdens of verification, error, and accountability onto lay users, leading to significant consequences for inequality and access to justice.

\section{Background and Theoretical Framework}
\label{sec:background}

\subsection{Access to Justice and the Infrastructural Deficit}

Access to justice has long been understood as comprising three interdependent components: legal knowledge, legal expertise, and access to appropriate legal fora \citep{lucy2020}. The first concerns citizens' capacity to recognize when their rights or obligations are engaged and to obtain reliable information about them \citep{wentz2005, sandefur2015b}. The second concerns the availability of qualified intermediaries capable of mediating between abstract legal rules and concrete factual circumstances \citep{sandefur2015a}. The third concerns the institutional pathways, including courts, tribunals, ombudsman schemes, and alternative dispute resolution bodies, through which legal claims are articulated and resolved \citep{prescott2017, koo2018, hannaford2003}.

All three have been substantially eroded over the last few years in jurisdictions that have pursued sustained curtailment of publicly funded legal services. For instance, in England and Wales, the Legal Aid, Sentencing and Punishment of Offenders Act (LASPO) 2012 significantly restricted the scope of civil legal aid, removing or limiting funding for many housing, welfare benefits, private family law, and non-asylum immigration matters, subject to important exceptions and exceptional case funding, producing what the Bach Commission \citep{hol2017} characterized as a crisis in access to justice and a threat to the rule of law. Subsequent empirical work has documented the consequences. Litigants in person constitute a substantial proportion of users in private family law proceedings, particularly following LASPO's removal of most private family cases from the scope of legal aid \citep{trinder2014}. The contraction has affected not only recipients but also the professionals who supply publicly funded legal services. In the years following LASPO, the number of civil legal aid provider offices completing work fell substantially, contributing to the emergence of what the Bach Commission \citep{hol2017} described as ``advice deserts'': areas with no advice centers, law centers or legal aid practices offering legally aided advice. This supply-side erosion is the predictable outcome of the neo-liberal restructuring of publicly funded legal services, a process that has reconfigured the position of both the recipients of legal aid and the practitioners who deliver it \citep{sommerlad2004}.

Funding cuts, however, are only part of the picture: even where legal aid exists, much justiciable need goes unrecognized or unpursued. Comparative legal needs research, drawing on the canonical contributions of \citet{genn1999}, \citet{pleasence2014}, and \citet{coumarelos2012}, has repeatedly shown that a substantial share of justiciable problems is never recognized as legal in nature, and that even when they are, professional advice is sought in only a minority of cases. Legal capability literature offers a complementary account, defining legal capability as the capacity to understand and act on justice problems and showing that its uneven distribution tracks and helps produce wider social inequalities \citep{pleasence2019}. Although the institutional form and severity of these deficits vary across jurisdictions, the underlying problem is not confined to England and Wales: in many legal systems, formal rights remain practically inaccessible where affordable advice, procedural support, and institutional assistance are unevenly distributed.

A similar pattern holds elsewhere. In the United States, the civil justice gap is well documented: the \citet{lsc2022} reports that 92\% of the civil legal problems of low-income Americans receive no or insufficient legal help. The consequences are not evenly distributed. As \citet{sandefur2019} argues, this should not be understood as a crisis of unmet legal need but as one of exclusion and inequality. Access is restricted, in that only some people and only some kinds of problems secure lawful resolution, and systematically unequal, in that wealthier and more advantaged groups obtain greater access than poorer and marginalized ones. On this account, the concern is not the supply of legal services but the just resolution of legal problems. This reframing widens the range of admissible solutions beyond the lawyer-centered model. This is the empirical setting in which the present analysis must be located. While the system of professionally mediated access to law was once expected to provide a baseline of legal protection, it has, in significant areas, ceased to function as usable infrastructure for those who depend on it. Under such conditions, alternative systems are repurposed to perform infrastructural roles and ensure access to justice. The analytic question is how, by whom, and with what consequences for the social production of legality.

\subsection{AI in Legal Contexts and the Limits of Accuracy as a Framing}

A growing literature presents artificial intelligence as a candidate response to each of the three dimensions of access to justice. AI is said to improve access to legal information through autonomous retrieval and simplified explanation \citep{tan2025}, to reduce the cost of legal assistance by automating routine correspondence and drafting \citep{doyle2025}, and to increase the efficiency of court administration \citep{chen2025}. Institutional actors have noticed this potential. The \citet{unesco2025} Guidelines for the Use of AI Systems in Courts and Tribunals, and the UK Ministry of Justice's AI Action Plan for Justice \citep{moj2025} frame AI as a partial answer to access deficits, while the Courts and Tribunals Judiciary's guidance to judicial office holders \citep{ctj2025} acknowledges the routine use of generative AI tools by litigants in person.

At the same time, critical literature has identified a series of structural problems that any account of AI in law must take seriously. The most visible of these is the risk of hallucinations, i.e., the production of plausible but fabricated content, including invented case citations \citep{munir2025, magesh2025}. The empirical reliability problem is real and well-documented. Nonetheless, this is an inadequate framing of what is at stake. Hallucination treats the problem as one of accuracy, as if a more reliable model would dissolve the difficulty. A more fundamental problem is that LLMs do not perform legal reasoning in the sense that legal practice requires. As \citet{hildebrandt2015} argues, legal reasoning in the constitutional tradition is tied to the interpretive and contestable character of legal text, and computational systems that operate on predictive rather than interpretive logics sit uneasily with the ends that law is meant to serve. As \citet{pasquale2020} argues, legal judgment draws on forms of contextual and relational expertise that resist automation. Furthermore, model performance on legal tasks tracks the prominence and availability of the underlying materials: large language models perform markedly better on highly cited, well-represented courts and jurisdictions than on less prominent ones, reflecting the uneven representation of legal sources in their training data \citep{dahl2024}.
These limitations bear on the present argument in a specific way. If LLMs do not, in any meaningful sense, perform legal reasoning, then the question of how their outputs come to be treated as legally credible is not technical but social. It is a question about how legal authority is attributed, contested, and stabilized under conditions in which the formal institutions that have historically performed that work is partially absent. This question cannot be answered by improving model accuracy. It requires a theoretical apparatus capable of describing how legality is attributed to everyday encounters with law, outside the institutional precincts of the legal profession.

\subsection{Lay Legal Practice and the Work of Recognition}

Two adjustments to the prevailing analytic vocabulary are necessary at this stage. The first is terminological. A binary distinction between ``lawyers'' and ``non-lawyers'' obscures more than it reveals. The legal profession is internally stratified across solicitors, barristers, paralegals, chartered legal executives, in-house counsel of varying qualifications, and the array of regulated and unregulated providers that have proliferated following the Legal Services Act 2007 \citep{sommerlad2015}. The category of ``non-lawyer'' is correspondingly heterogeneous, encompassing entirely unfamiliar laypersons, litigants in person of considerable procedural sophistication, unregulated advice providers, and professionals adjacent to but outside the regulated legal services market. The analysis that follows therefore adopts the more capacious term \emph{lay legal practice}, drawing on the legal needs and legal capability tradition \citep{pleasence2014, pleasence2019, pleasence2015, mcdonald2021}, to designate work undertaken outside any regulated legal advice relationship by lay users acting on their own legal matters.

The second adjustment is theoretical. The legal capability literature has revealed the knowledge, skills, and attributes required for a person to recognize and act on a legal problem, grounding these in Sen's capability approach and attending closely to their uneven social distribution \citep{pleasence2014, mcdonald2021, coumarelos2012}. A parallel literature on legal technology and access to justice has examined the delegation of legal tasks to software, whether through document automation, guided interviews, or, more recently, generative tools, and has debated its promise and its risks \citep{doyle2025, susskind2023}. What this work has largely envisaged, however, is delegation to purpose-built systems deployed or supervised within an institutional framework by courts, legal service providers, or regulated intermediaries. The configuration we examine is different in two respects. First, the system is a general-purpose commercial large language model used directly by the layperson outside any service relationship. Second, the credibility of its output is established not through professional or institutional accreditation but through validation by a lay platform community. It is this combination, ungoverned delegation paired with community-based credibility, that existing frameworks have not had occasion to theorize. To analyze it, we turn to the socio-legal scholarship on dispute transformation. \citet{felstiner1981}'s analysis distinguishes three operations through which an experience comes to be treated as a legal claim: \emph{naming} (recognizing an experience as injurious), \emph{blaming} (attributing responsibility to another), and \emph{claiming} (asserting a right against that party).

The framework is valuable because it treats naming, blaming, and claiming as practices capable of being performed or supported by actors other than the disputant. We do not claim that large language models possess agency in any strong sense, nor that they ``co-author'' legal claims as a human adviser would. Our claim is functional: the work of dispute transformation that Felstiner, Abel and Sarat located in the disputant, and that the legal capability literature locates in the capable individual, is in this configuration distributed across three elements: (a) the language model that drafts and articulates, (b) the lay user who directs and executes, and (c) the platform community that validates. Whether this distribution of function amounts to a distribution of agency, in the sense developed in actor-network and distributed-cognition scholarship, is a question the configuration raises but that our data cannot settle. What our data does show is that the function is no longer performed by, or reducible to, any single human actor.

The emphasis on legality as an everyday accomplishment is developed most influentially in \citet{ewick1998}, \emph{The Common Place of Law}, which conceptualizes legality as something ordinary people produce and reproduce through their encounters with law. Ewick and Silbey identify three orientations through which lay legal consciousness operates: \emph{before the law}, a deferential and formal orientation that treats law as a separate and majestic domain; \emph{with the law}, a strategic and instrumental orientation that treats law as a resource to be deployed; and \emph{against the law}, a resistant orientation that treats law as a system to be evaded or subverted. These orientations are not mutually exclusive but coexist and shift across situations, providing a means of examining whether users treat AI-generated legal materials deferentially, strategically, or resistively.

\subsection{The Infrastructural Conditions of Distributed Counsel}

The argument so far has established four things. The production of legal authority is a social accomplishment rather than a property intrinsic to legal texts; it has historically been mediated by the regulated legal profession; that mediation has, in significant areas, broken down under the fiscal and institutional pressures traced above; and the work of dispute transformation that this mediation performed, the naming, articulating, and lending of credibility to a legal claim, does not disappear when professional provision is limited or missing but is \emph{redistributed} through the hybrid three-element arrangement described above (LLM, lay-user, online community). The analytic task is therefore to explain how this hybrid arrangement distributes legal work and produces a provisional form of credibility. Infrastructure scholarship provides a broader account of the technological and institutional conditions within which that arrangement develops.

One resource is the infrastructure studies tradition initiated by \citet{star1994} and developed by \citet{star1999} and \citet{plantin2018}. Infrastructure on this account is not a stable object but a relational property of socio-technical arrangements. A system becomes infrastructural when it is embedded in routine practice, when it organizes subsequent action, when it operates transparently in use, and when it becomes visible primarily at moments of breakdown. \citet{plantin2018} extend this analysis to digital platforms, arguing that platforms increasingly occupy infrastructural positions in domains historically organized through public institutions. This raises the question of how platform-mediated arrangements compare with the public infrastructures they partially displace, particularly in respect of accountability and accessibility. Platform infrastructures do not merely host evaluation. They organize which accounts become visible, which responses are amplified, and which forms of expertise are legible. The production of credibility is therefore shaped by both platform architecture and community judgment. As we show in Section~\ref{sec:rq2}, this architectural shaping of visibility has direct epistemic consequences.

Another resource is \citet{cohen2019}'s account of the legal construction of informational capitalism. Cohen's central claim is not that the informational economy has outrun the law, but that law has been actively enlisted in building it. In this way, the platform has become the core organizing logic of economic life, and legal institutions have been reshaped in the process, often in ways that entrench the position of powerful information intermediaries. The relevance for the present argument is that legal authority and legal meaning are, on Cohen's account, increasingly produced through the practices and architectures of the informational economy rather than solely through the formal institutions of the legal system. The configuration we examine is a case in point as it is a general-purpose platform and its user community comes to perform functions that were previously the province of regulated professionals, and they do so through their own logics rather than those of the professional regime. Through this configuration, such guidance is interpreted, contested, and accorded credibility, and the norms through which this credibility is attributed differ markedly from those of the regulated legal profession. Regulatory intervention can itself displace practice onto further intermediaries whose privacy and accountability properties differ from those of the arrangement being regulated \citep{mehta2026}.

\subsection{Distributed Counsel and the Spectrum of Verification}

AI-assisted lay legal practice is the broader category within which the present analysis is situated. It refers to the use of general-purpose LLMs by lay users handling their own legal matters outside a regulated legal advice relationship. This category includes private reliance on an LLM, the use of AI-generated material in legal correspondence or proceedings, and cases in which users seek further evaluation from other people or sources. Narrowing this down, we define distributed counsel as occurring where AI-generated legal advice is subjected to observable third-party community evaluation as part of the user's decision-making or implementation process. We develop distributed counsel as an ideal type in \citet{weber1949}'s sense: an analytically clarified configuration whose defining elements are machine generation, user direction, and community evaluation. Its novelty lies not in any one element, but in their conjunction. Legal tasks have long been delegated to software, users have long sought legal help from online communities, and online communities in other professional domains have already been observed working out the terms on which AI use is acceptable \citep{yuce2026}. Against this background, distributed counsel describes the routing of AI-generated legal advice through lay community scrutiny.

Not all AI-assisted lay legal practice takes this form. Far more commonly, users act on machine-generated guidance without reporting any evaluation by a community, a legal professional, an authoritative source, or an independent system. Because this classification rests on what users disclose, it captures only the absence of reported verification; the figures below are accordingly an upper bound on unverified action. What makes such action possible is an epistemic constraint within the data, as Reddit users cannot assess whether AI-generated legal advice is legally correct, since that assessment requires the very expertise whose absence drove them to the tool. What they can assess is whether the output looks and functions like something that will work, its lawyer-like form, apparent specificity, consistency across models, or the emotional reassurance it provides. Action therefore proceeds on the basis of perceived credibility and practical usability rather than an independent determination of legal correctness. Community aggregation through platforms like Reddit may provide a reliability filter by surfacing more diagnostic evidence than individual judgment ordinarily supplies, although it remains bounded by the heuristic cues available to participants \citep{elmas2026}.

This configuration displays an infrastructural orientation in \citet{star1994}'s sense, embedded in routine problem-solving, organizing the steps that follow, and becoming visible chiefly when it breaks down through hallucination or error. The practices observed are predominantly consistent with \citet{ewick1998}'s ``with the law'' orientation, because users describe deploying legal language, documents, and procedures strategically to advance their own matters.

Three qualifications delimit the scope of distributed counsel. First, the concept is an ideal type, not a claim about the modal case. To say that legal credibility is attributed through distributed counsel is not to claim that the arrangement is equivalent to professional legal services, nor to say that it is to be welcomed or deplored; instead it is to identify a configuration and trace how it operates.

Second, the concept registers the persistence of structural asymmetry. The redistribution of legal work is described by users as making participation more manageable, but the accounts continue to reflect the \citet{galanter1974} advantage. Favorable-outcome narratives are most prevalent where the opposing party is also unrepresented and least prevalent where the opposing party has professional representation.

Finally, the dataset consists of 153 narrative accounts of LLM-assisted legal work and 5{,}341 community reactions. These are accounts of practice and support analysis of how users describe their engagement with these systems, the orientations they articulate, and the discursive process through which AI-generated guidance is treated as more or less credible within a community, and do not establish habituation in the strong, behavioral sense \citet{star1999} intends, which would require longitudinal observation of repeated use. What follows is therefore best understood as an account of the narrative emergence of an infrastructural orientation toward lay users using AI-generated legal advice evidenced in how users speak about these systems and how communities receive them.

\section{Methods}
\label{sec:methods}

\begin{figure*}[t]
  \centering
  \includegraphics[width=\textwidth]{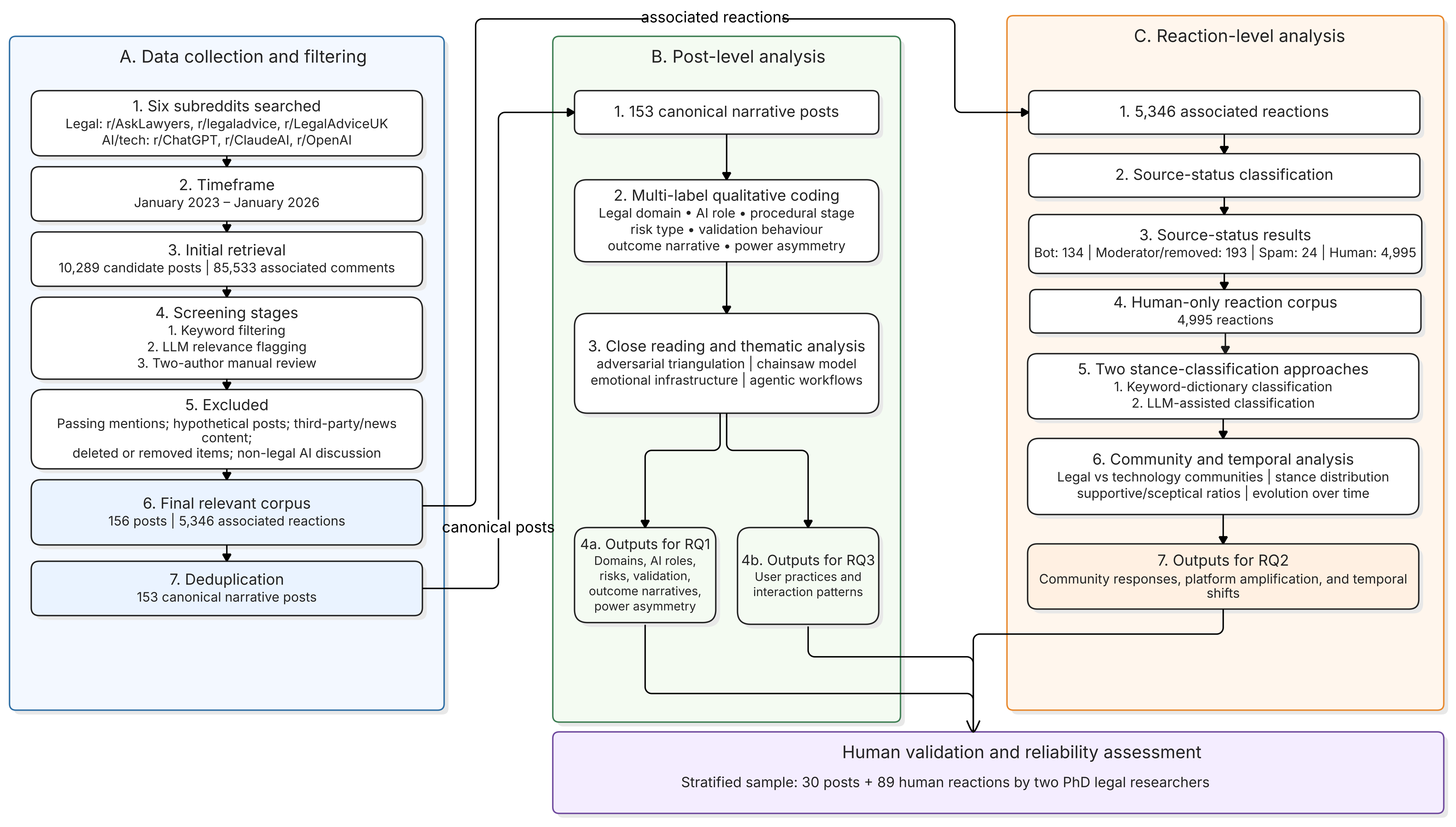}
  \caption{Three-stage analytical pipeline for collecting, coding, and analyzing Reddit discussions of AI use in legal contexts, with corresponding research-question outputs and human validation.}
  \label{fig:pipeline}
\end{figure*}

To undertake this study, we followed a three-stage procedure (see Figure~\ref{fig:pipeline}), described in detail below.

\subsection{Ethics}

This study received full ethical approval from the relevant institutional review board at the University of Edinburgh. All analyzed data were publicly posted on Reddit. Deleted posts were excluded, and no usernames, post identifiers, or other direct personal identifiers were retained.

Because the dataset includes sensitive legal disputes, findings are reported primarily in aggregate form. Distinctive narratives are paraphrased or omitted where quotation could create a risk of re-identification. Subreddit names are retained only where necessary for comparative analysis. Classification prompts are provided for transparency and reproducibility.

\subsection{Data Collection}

Using the API provided by Project Arctic Shift \citep{heitmann2026}, we collected posts and associated comments from six subreddits spanning legal and AI-tool communities, r/Ask\_Lawyers, r/legaladvice, r/LegalAdviceUK, r/ChatGPT, r/ClaudeAI, and r/OpenAI, covering January 2023 to January 2026. Applying a keyword filter (e.g., legal advice, lawyer, lawsuit, court) yielded an initial corpus of 10{,}289 candidate posts and 85{,}533 associated comments.

Screening then proceeded in three stages. First, the same keyword filter applied to the comments reduced the comment set to 377. Second, we used a LLM (Gemini 2.5 Flash) to flag posts and comments in which a user directly described using an AI tool to assist with a real legal matter, as opposed to merely mentioning AI or law in passing or posing a hypothetical. Third, two authors independently reviewed the flagged items against explicit inclusion criteria and resolved disagreements by discussion. Inclusion required a first-person account of actual, non-hypothetical LLM use on the author's own legal matter; we excluded passing mentions, third-party or news content, purely hypothetical or opinion posts, and posts that had been removed or deleted at the time of retrieval. This produced a final analytical corpus of 156 relevant posts. Near-duplicate posts (cross-posts and re-posts of the same underlying narrative) collapsed, removing 3 posts across 2 clusters, yielding 153 canonical narratives and 5{,}341 associated community reactions. All analyses use the 153 canonical narratives and their reactions.

\subsection{Analytical Approach}

Full coding definitions, keyword dictionaries, classification prompts, and denominator notes are provided in the Supplementary Material.

\subsubsection{Post-Level Coding and Qualitative Analysis}

The 153 canonical narrative posts were coded using a multi-label qualitative coding framework. Multi-label coding was necessary because individual posts often described more than one legal issue, more than one AI function, and more than one form of risk or validation. The coding framework captured seven dimensions: legal domain, AI role, procedural stage, risk type, validation behavior, outcome, and power asymmetry.

Legal domain codes identified the area of law or dispute involved, including debt, consumer and finance; housing and tenancy; contract, business, intellectual property and immigration; family, divorce, probate and estates; criminal and traffic matters; and employment and labor. AI role codes captured the function performed by the LLM, including drafting documents or letters, legal research and explanation, document review, strategy and negotiation support, emotional or cognitive scaffolding, and agentic or semi-autonomous workflow. Procedural stage codes distinguished pre-dispute assessment, active dispute or negotiation, litigation or filing, and post-judgment or enforcement.

Risk coding identified posts that described or evidenced reliability, safety, or governance concerns. These included policy refusal or safety-filter friction; hallucinated or incorrect law or citations; illegal or harmful guidance; privacy or surveillance concerns; and over-reliance without counsel. A binary ``any risk'' variable was also created. Validation coding distinguished two forms. Independent verification was captured as whether users corroborated AI output on their own, through cross-model comparison, or by checking it against authoritative legal sources or professionals; a binary ``any independent verification'' variable was created if any was present. Community validation was coded where a user presented AI-generated legal advice or material to a platform community and expressly or implicitly invited assessment, correction, confirmation, or critique. It was measured separately as part of the distributed counsel sequence (Section~\ref{sec:rq1}), in which retrospective comments that could not affect the user's conduct were not coded as distributed counsel.

Outcome coding captured self-reported outcome narratives, rather than independently verified legal outcomes. Posts were coded as favorable-outcome narratives where users claimed a concrete benefit, such as recovery of money, dismissal of a claim, successful settlement, avoided fees, successful filing, or another practical advantage attributed at least in part to AI use. Posts were coded as adverse-outcome narratives when users described procedural errors, reliance on false information, financial loss, escalation, emotional harm, or other legal disadvantages associated with AI use. These classifications therefore measure how users represent the consequences of LLM-assisted legal self-help on Reddit; they do not establish actual legal success, causation, or objective case outcomes.

For the 153 canonical narratives, we also coded power asymmetry. Power asymmetry refers to the structural relationship between the AI-using individual and the opposing party or legal context, including differences in institutional resources, legal expertise, financial capacity, procedural familiarity, and positional authority. Categories included individual versus corporate or institutional actor, individual versus represented individual, individual versus unrepresented individual, individual versus state, information-seeking with no adversary, and professional use of AI. This variable was used to examine whether the prevalence of favorable and adverse self-reported outcome narratives varied according to the user's structural position within the dispute.

In addition to structured coding, the posts were analyzed thematically through close reading. Thematic claims were interpreted alongside the coded counts, rather than treated as free-standing anecdotal evidence. The codebook was developed and applied by the authors through close reading of the corpus. Coding was multi-label and interpretive, and category boundaries were refined through discussion as coding proceeded. Because several dimensions involve interpretive judgments about situated accounts, we treat this coding as transparent qualitative coding grounded in an explicit codebook rather than as a measurement instrument; the full code definitions and decision rules are provided in the Supplementary Material to support scrutiny and reuse.

\subsection{Reaction Classification and Community Analysis}

Throughout, we use \emph{reaction} to denote a comment or reply (including nested replies) associated with one of the relevant posts, rather than a Reddit vote score; vote counts are not analyzed as reactions. The 153 posts generated 5{,}341 associated community reactions. These reactions were analyzed to examine how platform communities respond to accounts of AI-assisted legal self-help. We first classified reactions by source status using Gemini-2.5-flash. Each reaction was assigned to one of four categories: bot, moderator or removed, spam, or human. This identified 132 bot reactions, 192 moderator or removed reactions, 24 spam reactions, and 4{,}993 human reactions. Subsequent stance analysis was conducted on the human-only reaction corpus.

We then used two complementary stance-classification methods. First, we applied a keyword-dictionary method to classify explicit stance expressions. Seven stance categories were used: authenticity-questioning, dismissive or anti-AI, supportive or celebratory, skeptical or warning, lawyer or professional perspective, shared experience, and questions or curiosity. Reactions were matched against all dictionaries using case-insensitive substring matching. A reaction could receive multiple stance labels. This method is reproducible and conservative, but it under-detects sarcasm, implicit agreement, narrative responses, and context-dependent legal discussion.

Second, we used Gemini to conduct LLM-assisted stance classification of the 4{,}993 human reactions. Reactions were assigned to ten categories: tangential, substantive discussion, supportive, skeptical, neutral, shared experience, authenticity-questioning, curious, dismissive, and professional. The LLM-assisted method was used to capture context-dependent reactions that are difficult to identify through exact keyword matching. We treat this classification as a descriptive and interpretive aid rather than as ground truth and report it alongside the keyword-based results for methodological triangulation.

Both classification stages used the same model (Gemini-2.5-flash) under default decoding settings. Source-status and stance were assigned in two sequential passes, each returning a single dominant label per reaction; classification was single-pass and run in small batches. Each reaction was classified with its subreddit and the parent post's title supplied as context. Where a reaction expressed more than one stance, the most dominant was retained as the primary stance. Batches that failed to parse were assigned the conservative default at low confidence. The full prompts for both stages are reproduced in the Supplementary Material.

To examine platform differences, subreddits were grouped into two community types. Legal communities comprised r/Ask\_Lawyers, r/legaladvice, and r/LegalAdviceUK. Technology communities comprised r/ChatGPT, r/ClaudeAI, and r/OpenAI. We compared these community types by engagement volume, stance distribution, substantive discussion, authenticity-questioning, and supportive-to-skeptical ratios. Engagement was measured by the number of reactions per post. Amplification was assessed by comparing mean reactions per post across community types and by calculating the share of total reactions attributable to the most highly discussed posts.

\subsubsection{Human Validation and Inter-Annotator Agreement}

To assess the reliability of the coding scheme and the LLM-assisted classifications, a stratified validation sample of 30 posts and 89 associated human reactions was independently re-annotated by two doctoral researchers in law, working separately from the same codebook. The sample was stratified by community type, legal domain, and model-predicted stance. Agreement between the two researchers was substantial for the principal reaction-level variable, with Cohen's $\kappa = 0.66$ and Krippendorff's $\alpha = 0.66$ for the ten-category stance classification, alongside 98.9\% raw agreement for source status. For the multi-label post-level dimensions, mean pairwise Jaccard agreement was highest for validation behavior and legal domain, at 0.75 and 0.64 respectively, and lower for the more fine-grained risk typology, at 0.36, reflecting the interpretive character of situated legal coding. Against the adjudicated expert labels, the LLM-assisted stance classification achieved $\kappa = 0.74$. As supplementary measures, the model agreed with at least one researcher on 90\% of reactions for stance and 87\% of posts for legal domain. Full agreement statistics for all dimensions are reported in the Supplementary Material.

\subsubsection{Temporal and Cross-Tabular Analysis}

Posting dates were used to examine changes over time. The temporal analysis covered January 2023 to January 2026. For cohort-level comparisons, posts were grouped into three periods: early adoption, covering January to December 2023; middle adoption, covering January to December 2024; and late adoption, covering January 2025 to January 2026. These periods were used to compare legal domains, AI roles, risk types, validation behavior, procedural stage, subreddit distribution, and outcome reporting.

The terms ``quiet period'' and ``inflection point'' are used descriptively. The quiet period refers to November 2023 to March 2024, during which observed post volume was near zero. The inflection point refers to the subsequent period of sustained growth beginning around December 2024. These labels describe observed distributional patterns and do not imply causal attribution.

\section{Results}
\label{sec:results}

\subsection{RQ1: Pathways to Verification}
\label{sec:rq1}

Independent verification occurred in 17.3\% of the 153 canonical narratives. Separately, AI-generated legal advice was submitted for community evaluation in 30 of the 153 discussion threads (19.6\%). These threads were included only where community feedback could shape the user's decision, revision, or next step. The first measure reflects users' efforts to validate outputs, such as by consulting another model, an authoritative source, or a professional. The second measure involves scrutiny by a third-party platform community. As these measures pertain to distinct practices and employ different units of analysis, direct comparison is not appropriate.

In most instances, users reported neither form of verification. This classification rests on user disclosures and so captures the absence of reported verification; checking may have occurred offline. Where verification is reported, it takes three recurring forms.

\subsubsection{Human Oversight}

The AI tool-user relationship is most clearly articulated in a probate dispute case, where a user employs a metaphor to describe AI as a power tool, a versatile but dangerous tool that needs clear human oversight and careful management. In this post, the user is describing how the AI assists with discrete tasks while remaining dependent on user direction and control. They also foreground a key limitation of using LLMs for legal information: hallucinations, a concern echoed by users across the dataset. Posts consistently describe a division of labor in which LLMs handle drafting, research, and document assembly, while users retain responsibility for verification and procedural execution.

\subsubsection{Triangulation and Validation Across Models}

Fifteen posts (9.8\%) describe querying multiple LLMs in parallel to compare outputs, identify inconsistencies, and reduce the risk of hallucinations. Users report cross-checking responses across systems (e.g., ChatGPT and Gemini), with one describing this process as making the models ``argue with each other'' to identify errors and omissions, sometimes producing what is described as a ``full court ready case file.'' Rather than relying on a single output, these posts describe a practice of comparing responses across multiple systems to assess consistency.

This pattern is also combined with platform-mediated verification. Approximately 30 posts (19.6\%) describe a recurring sequence: AI query $\rightarrow$ AI response $\rightarrow$ Reddit post seeking community feedback. In these cases, users present AI-generated outputs to Reddit for evaluation, correction, or confirmation. Posts following this sequence frequently include requests for verification or critique of AI-produced material.

\subsubsection{AI as Emotional Infrastructure}

A distinct subset of posts (18; 11.8\%) describes AI systems that perform functions beyond providing legal information. Instead, they operate as a form of emotional infrastructure within AI-assisted lay legal practice. In these cases, AI is used not only to generate legal content and aid with tasks but also to sustain engagement over time through reminders, conversational continuity, and affective stabilization. For example, one user facing eviction describes how ChatGPT misclassified caregiver grants as income and, in every subsequent session, persistently prompts the user to contact a pro bono law firm by storing and reintroducing reminders.

Similarly, users navigating highly sensitive contexts, such as domestic violence, report using AI to produce ``emotionally neutral, legally useful messages,'' while maintaining a sense of control during periods of instability. These posts describe the use of AI to assist with drafting communications in situations involving ongoing conflict or distress. Across these cases, users report that AI helps them structure messages, manage tone, and respond consistently over time, particularly when direct communication is described as difficult or emotionally charged.

\subsection{RQ2: Distributed Counsel and Community Scrutiny}
\label{sec:rq2}

Reddit does not distribute attention evenly. Posts in technology forums receive many times the engagement of posts in legal forums, and a small number of threads dominate the conversation: the twenty most-discussed posts account for more than four-fifths of all reactions. The scrutiny a post receives is strongly associated with where it is posted.

\subsubsection{Platform Amplification}

The 5{,}341 reactions to the 153 posts are highly unevenly distributed across both communities and posts. Technology-oriented subreddits (\emph{r/ChatGPT}, \emph{r/ClaudeAI}, \emph{r/OpenAI}) generate substantially higher engagement than legal forums (\emph{r/legaladvice}, \emph{r/LegalAdviceUK}, \emph{r/Ask\_Lawyers}), with a mean of 38.4 reactions per post compared to 2.5, representing a 15.6-fold difference. This indicates that most interaction occurs in technology-focused communities rather than in legally oriented ones.

Engagement is also concentrated at the level of individual posts. The top 20 posts account for 82.5\% of all reactions, demonstrating a steeply skewed distribution in which a small subset of posts receives the majority of attention. As a result, patterns of discussion are disproportionately shaped by a limited number of highly visible cases rather than the dataset as a whole.

\subsubsection{LLM-Based Stance Classification}

LLM-based classification substantially increases analytical coverage relative to keyword matching (Table~\ref{tab:stance}). The most notable effect is the recovery of otherwise under-detected categories: \emph{shared experience} increases from 39 to 389 cases, \emph{supportive} reactions from 202 to 584, and \emph{curious} reactions from 31 to 224. Keyword methods, reliant on exact phrase matching, systematically under-identify positive and experiential engagement.

\begin{table}[t]
  \centering
  \caption{Stance Distribution}
  \label{tab:stance}
  \small
  \begin{tabular}{@{}lrr@{}}
    \toprule
    Stance & $N$ & \% \\
    \midrule
    Tangential              & 1{,}585 & 31.7 \\
    Substantive discussion  & 711     & 14.2 \\
    Supportive              & 583     & 11.7 \\
    Skeptical               & 427     & 8.5 \\
    Neutral                 & 418     & 8.4 \\
    Shared experience       & 389     & 7.8 \\
    Authenticity-questioning& 382     & 7.6 \\
    Curious                 & 224     & 4.5 \\
    Dismissive              & 197     & 3.9 \\
    Professional            & 77      & 1.5 \\
    \bottomrule
  \end{tabular}

  \vspace{3pt}
  {\footnotesize \emph{Note.} Each reaction was assigned a single dominant stance, so categories are mutually exclusive and percentages sum to 100. Percentages are of the 4{,}993 human reactions.}
\end{table}

\subsubsection{Polarized Community Norms}

Reaction profiles diverge systematically across community types (Table~\ref{tab:community}). Legal communities are dominated by substantive legal engagement (65.9\%), whereas technology communities exhibit a more diffuse and polarized distribution. The supportive-to-skeptical ratio diverges sharply: 1.4:1 in technology communities, compared to 0.15:1 in legal communities.

\begin{table}[t]
  \centering
  \caption{Stance by Community Type}
  \label{tab:community}
  \small
  \begin{tabular}{@{}lrrrr@{}}
    \toprule
    & \multicolumn{2}{c}{Legal} & \multicolumn{2}{c}{Tech} \\
    \cmidrule(lr){2-3}\cmidrule(lr){4-5}
    Stance & $N$ & \% & $N$ & \% \\
    \midrule
    Substantive discussion   & 176 & 65.9 & 535    & 11.3 \\
    Supportive               & 4   & 1.5  & 579    & 12.3 \\
    Skeptical                & 26  & 9.7  & 399    & 8.5  \\
    Shared experience        & 8   & 3.0  & 381    & 8.1  \\
    Authenticity-questioning & 1   & 0.4  & 381    & 8.1  \\
    Tangential               & 26  & 9.7  & 1{,}560 & 33.0 \\
    \bottomrule
  \end{tabular}

  \vspace{3pt}
  {\footnotesize \emph{Note.} Columns report the distribution of dominant stance within each community type; percentages are of all human reactions in that community. Only the six most analytically relevant stances are shown.}
\end{table}

Temporal analysis indicates a reversal in evaluative orientation. In 2023, the supportive-to-skeptical ratio was 0.77:1; by 2025--26, this had increased to 1.83:1. Over the same period, substantive discussion increased (11.5\% to 15.6\%), while neutral responses declined (11.8\% to 6.6\%), indicating a shift toward more evaluative and opinionated engagement.

\subsection{RQ3: The Landscape of AI Legal Self-Help}

AI-assisted lay legal practice is broad, consequential, and dominated by drafting. Post-level percentages below are based on the 153 canonical narratives; reaction-level percentages on the 4{,}993 human reactions.

\subsubsection{Legal Domains and Roles}

The 153 canonical posts span six legal domains, with activity concentrated in debt/consumer/finance and family/divorce/probate (Table~\ref{tab:domains}). This distribution indicates clustering of LLM-assisted legal activity within a limited set of recurrent problem types.

AI use is organized into six functional roles, reflecting consistent patterns in how users deploy LLMs across legal tasks. This included drafting of documents and letters (77.1\%) and evidence of the emerging practice of agentic automation (9.8\%), where AI operates semi-autonomously in multi-step legal workflows.

\begin{table}[t]
  \centering
  \caption{Legal Domain and AI Role Distribution}
  \label{tab:domains}
  \small
  \begin{tabular}{@{}lrr@{}}
    \toprule
    Legal domain & $n$ & \% \\
    \midrule
    Debt/consumer/finance    & 51 & 33.3 \\
    Family/divorce/probate   & 40 & 26.1 \\
    Housing/tenant           & 34 & 22.2 \\
    Contract/business/IP     & 33 & 21.6 \\
    Employment/labor         & 23 & 15.0 \\
    Criminal/traffic         & 20 & 13.1 \\
    \midrule
    \multicolumn{3}{@{}l}{\textit{AI role}}\\
    Drafting docs/letters    & 118 & 77.1 \\
    Legal research           & 62  & 40.5 \\
    Strategy/negotiation     & 50  & 32.7 \\
    Document review          & 49  & 32.0 \\
    Emotional support        & 18  & 11.8 \\
    Agentic automation       & 15  & 9.8  \\
    \bottomrule
  \end{tabular}

  \vspace{3pt}
  {\footnotesize \emph{Note.} Categories are not mutually exclusive; posts could receive more than one code, so percentages sum to more than 100. Percentages are of the 153 canonical narratives.}
\end{table}

The majority of posts (77\%) involve AI drafting documents, including demand letters, motions to dismiss, court filings, regulatory complaints, settlement agreements, and General Data Protection Regulation (GDPR) complaints. The predominance of drafting suggests that, in these self-reported accounts, users often treat LLMs as practical tools for producing legally structured documents rather than solely for general information-seeking. Indeed, the quality of AI-generated drafts is generally described as serviceable to impressive, with several posts noting that the structure and tone of a professionally worded letter are the primary value delivered. Several users explicitly note that the formality of a ``lawyer-style'' letter, even if imperfect, caused landlords or employers to take them seriously for the first time. In one case, a user reported that a landlord waived a fee a few minutes after receiving a legally structured letter citing relevant law. That the form of a legal document, its structure, tone, and register, can produce effects seemingly independent of the soundness of its underlying claims is a pattern we return to in Section~\ref{sec:discussion}, where we read it as the social force of juridical language rather than of law itself.

Posts narrate that AI use is often combined within single problem-solving trajectories, with multiple functions appearing within the same case. These include legal research (information retrieval), drafting documents and correspondence, and instances in which users describe AI used across multiple steps in a process. In particular, 15 posts (9.8\%) describe uses consistent with agentic automation, where AI is used to support multi-step workflows rather than a single discrete task.

\subsubsection{Tools Used and the Evolving Landscape}

Tool usage is highly concentrated, with ChatGPT appearing in 90.2\% of posts, Claude appearing in 16 posts (10.5\%), Gemini/Bard in 11 (7.2\%), and Grok in 5. The temporal distribution exhibits rapid expansion consistent with infrastructural uptake dynamics, increasing from 17 posts in 2023 to 89 in 2025, with January 2026 producing 16 posts.

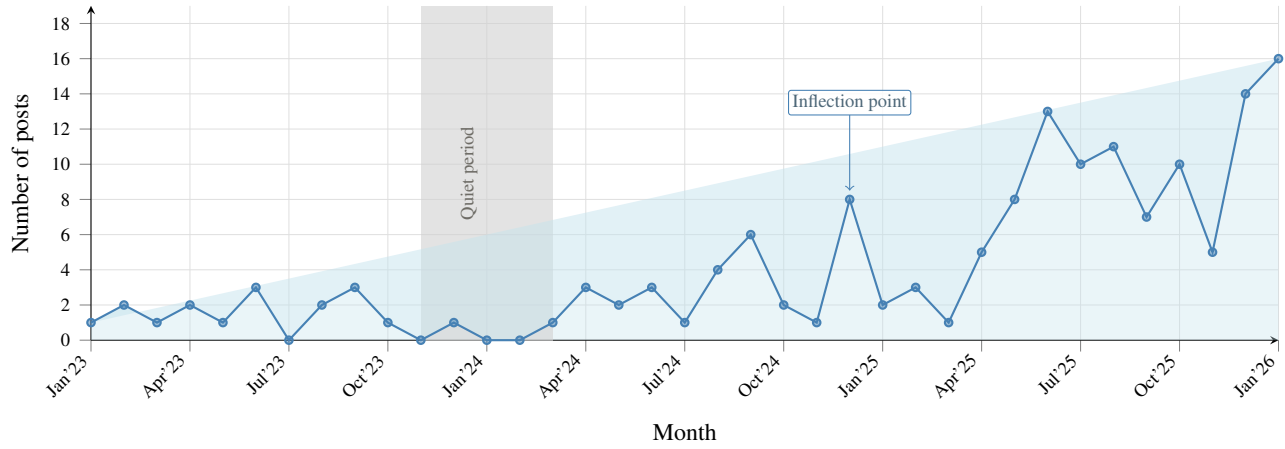
\begin{figure*}[t]
\centering
\begin{tikzpicture}
\begin{axis}[
    width=\textwidth,     
    height=6cm,
    xlabel={Month},
    ylabel={Number of posts},
    xmin=0, xmax=36,
    ymin=0, ymax=19,
    xtick={0,3,6,9,12,15,18,21,24,27,30,33,36},
    xticklabels={%
        {Jan'23},{Apr'23},{Jul'23},{Oct'23},%
        {Jan'24},{Apr'24},{Jul'24},{Oct'24},%
        {Jan'25},{Apr'25},{Jul'25},{Oct'25},%
        {Jan'26}},
    xticklabel style={rotate=45, anchor=east, font=\scriptsize},
    ytick={0,2,4,6,8,10,12,14,16,18},
    yticklabel style={font=\scriptsize},
    ylabel style={font=\small},
    xlabel style={font=\small},
    grid=major,
    grid style={gray!25, very thin},
    axis lines=left,
    every axis plot/.append style={thick},
    clip=false,
    tick align=outside,
    minor tick num=0,
]

\fill[quietshade, opacity=0.6]
    (axis cs:10,0) rectangle (axis cs:14,19);
\node[font=\scriptsize, text=warmgrey, rotate=90, anchor=south]
    at (axis cs:12,9.5) {Quiet period};

\addplot[
    name path=curve,
    fill=lightfill,
    fill opacity=0.5,
    draw=steelblue,
    thick,
    mark=*,
    mark size=1.4pt,
    mark options={fill=steelblue, draw=steelblue},
] coordinates {
    (0,1)   (1,2)   (2,1)   (3,2)   (4,1)   (5,3)
    (6,0)   (7,2)   (8,3)   (9,1)   (10,0)  (11,1)
    (12,0)  (13,0)  (14,1)  (15,3)  (16,2)  (17,3)
    (18,1)  (19,4)  (20,6)  (21,2)  (22,1)  (23,8)
    (24,2)  (25,3)  (26,1)  (27,5)  (28,8)  (29,13)
    (30,10) (31,11) (32,7)  (33,10) (34,5)  (35,14)
    (36,16)
};

\addplot[name path=baseline, draw=none, forget plot]
    coordinates {(0,0)(36,0)};
\addplot[lightfill, opacity=0.35, forget plot]
    fill between[of=curve and baseline];

\node[font=\scriptsize, text=darkbluegrey,
      fill=white, fill opacity=0.85,
      inner sep=1.5pt, draw=steelblue, thin, rounded corners=1pt]
    at (axis cs:23,13.5) {Inflection point};
\draw[->, steelblue, thin]
    (axis cs:23,12.8) -- (axis cs:23,8.5);

\end{axis}
\end{tikzpicture}
\caption{Temporal distribution of Reddit posts describing AI use for legal
issues, from January~2023 to January~2026. The shaded region marks a
quiet period (November~2023--March~2024) with near-zero activity. An
inflection point at December~2024 marks the beginning of sustained growth,
with monthly counts rising from single digits to 10--16 posts per month.}
\label{fig:temporal}
\end{figure*}

Temporal distribution of observed post volume of Reddit posts describing AI use for legal issues, from January 2023 to January 2026 (Figure~\ref{fig:temporal}). The shaded region marks a quiet period (November 2023--March 2024) with near-zero activity. An inflection point at December 2024 marks the beginning of sustained growth, with monthly counts rising from single digits to 10--16 posts per month.

This temporal pattern is accompanied by two notable shifts (Figure~\ref{fig:temporal}):

\begin{enumerate}[leftmargin=1.4em,itemsep=2pt,topsep=3pt]
  \item \textbf{Domain expansion}: Family/divorce/probate grew from 0\% (2023) to 23.8\% (2025--26).
  \item \textbf{Risk evolution}: Safety filter friction rose from 5.9\% (2023) to 21.9\% (2025--26), while overreliance without counsel rose from 0\% to 9.5\%.
\end{enumerate}

\subsubsection{Outcomes Narratives Associated with LLM Use}

Throughout this section, ``favorable'' and ``adverse'' refer to how users narrate and attribute the consequences of LLM use, not to verified legal results. Among posts describing a determinate consequence, favorable narratives outnumber adverse ones by roughly 20 to 1: just over two-thirds report a favorable outcome, while clear adverse outcomes appear in a handful of cases. These figures should be read as evidence of how users narrate and attribute outcomes in platform discourse, not as verified measures of legal effectiveness.

Turning to the distribution of power asymmetry, AI legal self-help occurs predominantly in asymmetrical disputes involving lay users and institutional actors: 44.4\% of disputes involve lay users facing corporations or institutions (Table~\ref{tab:power}).

\begin{table}[t]
  \centering
  \caption{Outcome Narratives by Power Asymmetry}
  \label{tab:power}
  \small
  \begin{tabular}{@{}lrrr@{}}
    \toprule
    Power configuration & $n$ & Fav.\ \% & Adv.\ \% \\
    \midrule
    Indiv.\ vs.\ corporate/inst.       & 68 & 45.6 & 2.9 \\
    Indiv.\ vs.\ unrepresented indiv.  & 18 & 50.0 & 0.0 \\
    Indiv.\ vs.\ represented indiv.    & 31 & 25.8 & 6.5 \\
    Indiv.\ vs.\ state                 & 16 & 31.3 & 6.3 \\
    Info-seeking, no adversary         & 12 & ---  & --- \\
    Professional using AI              & 8  & ---  & --- \\
    \bottomrule
  \end{tabular}

  \vspace{3pt}
  {\footnotesize \emph{Note.} Fav.\ \% and Adv.\ \% are row-wise: the percentage of posts within each power configuration that narrate a favorable or adverse outcome. Posts without a determinate outcome are omitted.}
\end{table}

Outcome differentials track power configuration, with self-reported favorable-outcome rates decreasing as opposing parties exhibit greater formal legal representation. Narratives of successful resolution are most frequent in disputes users describe as between unrepresented parties (50.0\%) and least frequent in disputes users describe as involving a represented opponent (25.8\%). This patterning of accounts echoes the asymmetry \citet{galanter1974} identifies. The prevalence of outcome narratives also varies by domain: housing/tenancy posts most often narrate a favorable outcome (44.1\%), whereas family/divorce/probate posts do so least often (17.2\%) and most often narrate harm (6.9\%).

Risk and verification are unevenly matched across the corpus. Nearly half of all posts (46.8\%) carry at least one documented risk code, yet only 17.3\% show independent verification. Common risks are policy refusal and safety-filter friction (19.6\%), hallucinated or incorrect citations (13.1\%), and illegal or harmful guidance (11.1\%). These risks are associated with different reported outcome profiles: posts describing hallucinated or incorrect legal information contain the lowest proportion of favorable-outcome narratives and the highest proportion of adverse-outcome narratives among the risk categories examined (15.0\% favorable, 10.0\% adverse). For example, one lawyer reports that GPT-4 fabricated 9 out of 10 case citations in a motion; in another case, a user nearly failed to serve a defendant after relying on a hallucinated court address.

\section{Discussion}
\label{sec:discussion}

The Reddit users examined here mostly do not verify AI-generated legal advice. When verification occurs, it takes the form of adversarial triangulation across models and, in the fullest configuration, distributed counsel, which is the submission of machine-generated guidance to online communities for scrutiny. In the majority pattern, AI-generated legal advice is acted on because it reads as lawyer-like, specific, and reassuring. This takes the form of a judgment about sufficiency for action, not correctness.

It is within this uneven distribution of work and responsibility that the infrastructural role of LLMs comes into view. Under conditions of constrained access to legal assistance, users describe folding these systems into the everyday tasks of drafting, research, document review, and preliminary strategic assessment, while retaining responsibility for verification and procedural execution. The predominance of drafting and legal research indicates that users generally approach law as a resource to be mobilized. AI-generated letters, filings, and legal formulations are valued principally for what they enable users to do within disputes, rather than as objects of abstract legal understanding. This pattern is consistent with \citet{ewick1998}'s with the law orientation.

We also found that to manage the guidance these systems produce, users develop practical strategies of their own, including adversarial triangulation, comparison against community advice, and iterative prompting aimed at surfacing errors or inconsistencies. In \citet{star1999}'s terms, they begin to acquire the situated knowledge required to operate within the arrangement, even if the present data cannot establish routinization or dependency in a strong longitudinal sense.

Alongside this practical role, LLMs also function as emotional infrastructure. Their capacity for continuous, responsive, and non-judgmental interaction may help users manage uncertainty and distress, particularly in high-stakes contexts. This affective function may be constitutive of how some users sustain engagement with AI-assisted lay legal practice. By offering reassurance during stressful legal processes, AI may help some users continue engaging with legal problems and related tasks. The narratives also suggest that affective support can coexist with heightened trust in the guidance, although the present data cannot establish whether emotional reassurance causes reduced scrutiny or increased reliance.

The infrastructural character of these systems also becomes visible at moments of breakdown \citep{star1999}, when outputs are inaccurate, incomplete, or contested, and users turn to platform communities to repair, validate, or reject AI-generated legal advice. In this context, breakdown manifests as hallucinations, inconsistencies, or failures to account for jurisdictional specificity, prompting users to reassess the reliability of LLM-generated outputs. These moments render the otherwise taken-for-granted role of LLMs perceptible, exposing their underlying limitations and the distributed nature of verification practices.

These findings extend access-to-justice scholarship by showing how everyday users access AI-generated legal advice in the retreat from professional legal services. We report that users consistently describe AI systems as providing immediate, low-cost, and iteratively responsive interactional environments. These affordances enable multi-turn dialogue through which legal understanding is progressively refined and indicate repeated, iterative forms of engagement across diverse legal problem types. The temporal shift in community responses, from a supportive-to-skeptical ratio of 0.77:1 in 2023 to 1.83:1 in 2025--26, indicates a marked reorientation in platform-level discourse. The temporal shift is consistent with increasing acceptance of AI-assisted legal self-help within the observed Reddit discussions. We also show that, in this context, Reddit is a functional component of this emerging infrastructure, shaping the visibility, credibility, and circulation of AI-mediated legal strategies and providing a layer of friction for verifying responses. However, as LLMs become more advanced, it remains unclear whether users will continue to rely on this intermediary layer of human validation or instead default directly to LLM-generated answers.

Our results also show that this infrastructural integration is not epistemically neutral, and that there is a sharp imbalance between technology and legal-community reactions to LLM-generated legal answers. Because visibility determines which interpretations users are most likely to encounter, technology-community responses, where supportive views predominate (supportive:skeptical $= 1.4$:1), are disproportionately amplified. In contrast, legal communities engage primarily through substantive legal analysis (65.9\%), with skepticism outweighing support (supportive:skeptical $= 0.15$:1), yet these perspectives reach a comparatively limited audience. As a result, interpretations shaped by technological optimism become more visible, while legally grounded evaluation is comparatively marginalized online.

This asymmetry is further reinforced by divergent credibility norms within the communities. Technology-oriented communities frequently interrogate the authenticity of posts (382 instances), reflecting concerns about AI-generated content and commercial promotion. Legal communities, by contrast, rarely engage in authenticity questioning (1 instance) and instead focus on the substantive legal dimensions of the issues presented. While concerns about astroturfing are empirically justified, including the identification of spam posts, this orientation toward authenticity may inadvertently displace attention from assessing legal accuracy. As a result, within technology-oriented spaces, evaluative attention is often directed toward whether a claim is genuine rather than whether it is legally correct. This also accounts for an apparent tension in our data. Substantive legal evaluation is concentrated in low-visibility legal forums, where 65.9\% of reactions involve substantive discussion but posts receive only 2.5 reactions on average. Technology forums generate far greater visibility, averaging 38.4 reactions per post, but contain more authenticity-questioning and a substantially lower proportion of substantive legal discussion.

This decoupling of credibility from correctness has a counterpart in the way AI-generated documents act on their addressees. Repeated reports that a lawyer-like letter can pressure a landlord or employer into conceding illustrate the force not of law but of its appearance: juridical language can produce real effects through its form and air of authority alone, even when the claims it advances are unsound \citep{bourdieu1987}.

This redistribution of legal work alters the conditions of participation without disturbing the hierarchy of outcomes. Our power-asymmetry data tracks the pattern predicted by \citet{galanter1974}: self-reported favorable outcomes are most common when the opposing party is also unrepresented (50.0\%) and least common when the opposing party is represented (25.8\%).

The expansion of AI-generated legal advice into more complex legal domains, including family law and disputes requiring sustained multi-party engagement, introduces additional risks. Regulatory frameworks governing the provision of legal advice, particularly rules concerning the unauthorized practice of law, remain jurisdictionally fragmented, and the legal status of advice generated by LLMs, as well as the allocation of liability among providers, remains uncertain \citep{gupta2025, cheong2024}. These uncertainties are compounded by empirical indicators of reliability risk. Across the dataset, only 17.3\% of posts report independently verifying AI outputs, while 46.8\% of posts contain identifiable risk markers, and hallucinated or incorrect law or citations are documented in 13.1\% of posts. This combination of low verification, high reliance, and documented error rates creates conditions for miscalibrated trust. In turn, this can raise questions about accountability in contexts where LLMs increasingly function as infrastructural intermediaries for legal advice.

Systematic validation remains rare. Users have improvised strategies like adversarial triangulation across models, chief among them, but these are exceptions within a practice that mostly proceeds unchecked. This pattern aligns with established literature on automation bias, namely the tendency to over-rely on AI systems \citep{strauss2021}, a phenomenon also documented among legal professionals \citep{kluttz2019}. Accordingly, there is a need to ensure that users are made aware of the limitations of LLM-generated outputs, including their susceptibility to hallucination, inconsistency, and context sensitivity, and are supported in adopting verification practices that preserve independent judgment. The risk of automation bias is particularly acute in contexts where AI functions as an emotional infrastructure, for example, in situations involving domestic violence, where outputs may be afforded heightened trust, and incorrect advice may have serious consequences.

It should be acknowledged that this study has several limitations. First, the analysis relies on self-reported Reddit posts, which may involve inaccuracies, selective disclosure, and unverified legal outcomes. Posters may be more likely to share successful or unusual cases, and platform amplification may further privilege favorable or dramatic narratives. The findings, therefore, cannot establish objective legal effectiveness, causation, or actual case success. Moreover, because the corpus captures only users who chose to narrate their experiences publicly, any wholly private reliance on AI-generated advice is, by construction, invisible. Accordingly, the observed proportion of narratives reporting verification should be treated as a lower-bound estimate of actual verification, while the proportion classified as lacking reported verification represents an upper-bound estimate of genuinely unverified reliance.

However, this does not undermine the central purpose of the study, which is to examine how users come to treat AI-generated legal advice as credible enough to act upon. Although the sample is limited, 153 posts and 5{,}341 reactions provide a sufficiently substantial corpus to identify recurrent patterns of use, validation, risk, and community response to AI-generated legal information. Second, commercial manipulation cannot be excluded. Some posts may reflect spam, astroturfing, or promotional activity, which may inflate the number of positive accounts of AI legal use. This risk is partly addressed by separating bot, spam, removed, and human reactions, and by treating authenticity-questioning as part of the platform dynamics under analysis. Third, the use of Gemini for reaction classification introduces both a risk of model bias and a reflexive tension, since our pipeline reproduces, in miniature, the configuration we study: a machine-generated output subjected to human validation. This risk is mitigated by triangulating the classifications with a reproducible keyword method and benchmarking them against independent expert annotations. Finally, the temporal findings cannot establish why attitudes changed over time.

\section{Conclusion}
\label{sec:conclusion}

This paper asks whether, how, and by whom AI-generated legal advice is verified before lay users act on it. Mostly, it is not. We report that a minority of users verify through cross-model triangulation or, in the fullest form, distributed counsel, in which an LLM generates advice, a lay user directs and applies it, and a platform community evaluates it. In the majority pattern, however, users act without any verification reported, and the burden of deciding whether machine-generated guidance is good enough rests solely on them.

Its implications for access to justice are complex. First, access should not be assessed solely by whether legal advice is available, but by the conditions under which that advice acquires credibility and can be translated into effective action. Second, legal authority can no longer be understood as flowing exclusively from professional status or formal institutions. In AI-mediated legal self-help, credibility is assembled from the form of the guidance, not from checks upon it. The absence of verification from most accounts is itself the finding. The central challenge is not whether AI-generated legal advice is accurate. It is how to govern an arrangement in which such advice becomes actionable without the safeguards of regulated legal services.

Future research should pursue two related directions. The first is comparative and longitudinal, examining whether distributed counsel develops differently across platforms such as TikTok, X, and specialized legal forums, and how platform affordances and community norms shape the credibility of AI-generated guidance. The second concerns the emotional dimension of AI-generated legal advice. LLMs can help users manage distress and sustain engagement with legal problems. However, the same trust that supports perseverance may also reduce scrutiny of inaccurate or harmful advice. This tension warrants closer examination, particularly in high-stakes and vulnerable contexts in which AI-generated legal advice is used.

\newpage
\bibliographystyle{apalike}
\bibliography{ref}

@article{bourdieu1987,
  author = {Bourdieu, Pierre},
  title = {The Force of Law: Toward a Sociology of the Juridical Field},
  journal = {Hastings Law Journal},
  volume = {38},
  number = {5},
  pages = {814},
  year = {1987}
}

@article{chaserant2013,
  author = {Chaserant, Camille and Harnay, Sophie},
  title = {The Regulation of Quality in the Market for Legal Services: Taking the Heterogeneity of Legal Services Seriously},
  journal = {European Journal of Comparative Economics},
  volume = {10},
  number = {2},
  pages = {267--291},
  year = {2013}
}

@article{chen2025,
  author = {Chen, Qingxia},
  title = {Improving the Trial Efficiency of Criminal Cases with the Assistance of Artificial Intelligence},
  journal = {Discover Artificial Intelligence},
  volume = {5},
  number = {1},
  pages = {110},
  year = {2025}
}

@inproceedings{cheong2024,
  author = {Cheong, Inyoung and Xia, King and Feng, KJ Kevin and Chen, Quan Ze and Zhang, Amy X.},
  title = {(A) I Am Not a Lawyer, but...: Engaging Legal Experts Towards Responsible LLM Policies for Legal Advice},
  booktitle = {Proceedings of the 2024 ACM Conference on Fairness, Accountability, and Transparency},
  pages = {2454--2469},
  year = {2024}
}

@book{cohen2019,
  author = {Cohen, Julie E.},
  title = {Between Truth and Power: The Legal Constructions of Informational Capitalism},
  publisher = {Oxford University Press},
  year = {2019}
}

@article{cohen2024,
  author = {Cohen, Julie E.},
  title = {Platforms, Data Infrastructures, and Infrastructure Stacks},
  journal = {Global Governance by Data: Infrastructures of Algorithmic Rule, Georgetown University Law Center Research Paper},
  number = {2023/25},
  year = {2024}
}

@techreport{coumarelos2012,
  author = {Coumarelos, C. and Macourt, D. and People, J. and McDonald, H. M. and Wei, Z. and Iriana, R. and Ramsey, S.},
  title = {Legal Australia-Wide Survey: Legal Need in Australia},
  institution = {Law and Justice Foundation of New South Wales},
  year = {2012}
}

@techreport{ctj2025,
  author = {{Courts and Tribunals Judiciary}},
  title = {Artificial Intelligence (AI) Guidance for Judicial Office Holders},
  year = {2025},
  note = {Accessed July 3, 2026}
}

@article{dahl2024,
  author = {Dahl, Matthew and Magesh, Varun and Suzgun, Mirac and Ho, Daniel E.},
  title = {Large Legal Fictions: Profiling Legal Hallucinations in Large Language Models},
  journal = {Journal of Legal Analysis},
  volume = {16},
  number = {1},
  pages = {64--93},
  year = {2024}
}

@article{doyle2025,
  author = {Doyle, Colin},
  title = {Automation and Access to Justice},
  journal = {American Journal of Law and Equality},
  volume = {5},
  pages = {48--88},
  year = {2025}
}

@article{elmas2026,
  author = {Elmas, Tuğrulcan},
  title = {Humans Cannot Detect {AI}-Generated Media But Communities May---For Now: Collaborative {AI} Detection in {r/RealOrAI} on {Reddit}},
  journal = {arXiv preprint arXiv:2605.24287},
  year = {2026}
}

@book{ewick1998,
  author = {Ewick, Patricia and Silbey, Susan S.},
  title = {The Common Place of Law: Stories from Everyday Life},
  publisher = {University of Chicago Press},
  year = {1998}
}

@article{felstiner1981,
  author = {Felstiner, William L. F. and Abel, Richard L. and Sarat, Austin},
  title = {The Emergence and Transformation of Disputes: Naming, Blaming, Claiming...},
  journal = {Law \& Society Review},
  volume = {15},
  number = {3--4},
  pages = {631--654},
  year = {1981}
}

@article{galanter1974,
  author = {Galanter, Marc},
  title = {Why the `Haves' Come Out Ahead: Speculations on the Limits of Legal Change},
  journal = {Law \& Society Review},
  volume = {9},
  number = {1},
  pages = {95--160},
  year = {1974}
}

@book{genn1999,
  author = {Genn, Hazel},
  title = {Paths to Justice: What People Do and Think about Going to Law},
  publisher = {Hart Publishing},
  year = {1999}
}

@article{gupta2025,
  author = {Gupta, Suvrajyoti},
  title = {Open the UPL Gates and Let the Robot-Lawyers Walk In},
  journal = {Asian Journal of Legal Education},
  volume = {12},
  number = {2},
  pages = {138--152},
  year = {2025}
}

@article{hadfield2014,
  author = {Hadfield, Gillian K.},
  title = {The Cost of Law: Promoting Access to Justice Through the (Un)Corporate Practice of Law},
  journal = {International Review of Law and Economics},
  volume = {38},
  pages = {43--63},
  year = {2014}
}

@article{hannaford2003,
  author = {Hannaford-Agor, Paula and Mott, Nicole},
  title = {Research on Self-Represented Litigation: Preliminary Results and Methodological Considerations},
  journal = {Justice System Journal},
  volume = {24},
  number = {2},
  pages = {163--181},
  year = {2003}
}

@misc{heitmann2026,
  author = {Heitmann, Arthur},
  title = {Project Arctic Shift},
  howpublished = {GitHub repository},
  year = {2026},
  note = {Accessed July 3, 2026}
}

@book{hildebrandt2015,
  author = {Hildebrandt, Mireille},
  title = {Smart Technologies and the End(s) of Law: Novel Entanglements of Law and Technology},
  publisher = {Edward Elgar},
  year = {2015}
}

@techreport{hol2017,
  author = {{House of Lords Library}},
  title = {Bach Commission Report: The Right to Justice},
  institution = {House of Lords},
  number = {Research Briefing LLN-2017-0093},
  year = {2017}
}

@article{kant2024,
  author = {Kant, Manuj and Kant, Manav and Nabi, Marzieh and Carlson, Preston and Ma, Megan},
  title = {Equitable Access to Justice: Logical LLMs Show Promise},
  journal = {arXiv preprint arXiv:2410.09904},
  year = {2024}
}

@article{kluttz2019,
  author = {Kluttz, Daniel N. and Mulligan, Deirdre K.},
  title = {Automated Decision Support Technologies and the Legal Profession},
  journal = {Berkeley Technology Law Journal},
  volume = {34},
  number = {3},
  pages = {853--890},
  year = {2019}
}

@article{koo2018,
  author = {Koo, Anna K. C.},
  title = {The Role of the English Courts in Alternative Dispute Resolution},
  journal = {Legal Studies},
  volume = {38},
  number = {4},
  pages = {666--683},
  year = {2018}
}

@techreport{lsc2022,
  author = {{Legal Services Corporation}},
  title = {The Justice Gap: The Unmet Civil Legal Needs of Low-Income Americans},
  institution = {Legal Services Corporation},
  year = {2022}
}

@article{lorek2024,
  author = {Lorek, Laura A.},
  title = {AI Legal Innovations: The Benefits and Drawbacks of ChatGPT and Generative AI in the Legal Industry},
  journal = {Ohio Northern University Law Review},
  volume = {50},
  number = {3},
  pages = {4},
  year = {2024}
}

@article{lucy2020,
  author = {Lucy, William},
  title = {Access to Justice and the Rule of Law},
  journal = {Oxford Journal of Legal Studies},
  volume = {40},
  number = {2},
  pages = {377--402},
  year = {2020}
}

@article{magesh2025,
  author = {Magesh, Varun and Surani, Faiz and Dahl, Matthew and Suzgun, Mirac and Manning, Christopher D. and Ho, Daniel E.},
  title = {Hallucination-Free? Assessing the Reliability of Leading AI Legal Research Tools},
  journal = {Journal of Empirical Legal Studies},
  volume = {22},
  number = {2},
  pages = {216--242},
  year = {2025}
}

@article{mcdonald2021,
  author = {McDonald, Hugh},
  title = {Assessing Access to Justice: How Much `Legal' Do People Need and How Can We Know?},
  journal = {UC Irvine Law Review},
  volume = {11},
  pages = {693},
  year = {2021}
}

@article{mehta2026,
  author = {Mehta, Dhyey and Jalilzade, Eldar and Kalameyets, Maksim and Owens, Rebecca and Juarez, Marc and Aidinlis, Stergios and Shi, Lei and Elmas, Tuğrulcan},
  title = {Online Safety Regulation Increases Privacy Risk: Evidence from the {UK} {O}nline {S}afety {A}ct},
  journal = {arXiv preprint arXiv:2606.05273},
  year = {2026}
}

@techreport{moj2025,
  author = {{Ministry of Justice}},
  title = {AI Action Plan for Justice},
  institution = {Ministry of Justice},
  year = {2025},
  note = {Accessed July 3, 2026}
}

@article{munir2025,
  author = {Munir, Bakht},
  title = {Hallucinations in Legal Practice: A Comparative Case Law Analysis},
  journal = {International Journal of Law, Ethics and Technology},
  pages = {126},
  year = {2025}
}

@article{nielsen2024,
  author = {Nielsen, Aileen and Skylaki, Stavroula and Norkute, Milda and Stremitzer, Alexander},
  title = {Building a Better Lawyer: Experimental Evidence That Artificial Intelligence Can Increase Legal Work Efficiency},
  journal = {Journal of Empirical Legal Studies},
  volume = {21},
  number = {4},
  pages = {979--1022},
  year = {2024}
}

@inproceedings{pandit2026,
  author = {Pandit, Harshvardhan J. and Blankvoort, Dick A. H. and Shaaban, Adel and Luccioni, Sasha and Birhane, Abeba},
  title = {Terms of (Ab)Use: An Analysis of GenAI Services},
  booktitle = {The 2026 ACM Conference on Fairness, Accountability, and Transparency},
  pages = {5043--5067},
  year = {2026}
}

@book{pasquale2020,
  author = {Pasquale, Frank},
  title = {New Laws of Robotics: Defending Human Expertise in the Age of AI},
  publisher = {The Belknap Press of Harvard University Press},
  year = {2020}
}

@article{perlman2023,
  author = {Perlman, Andrew},
  title = {The Implications of ChatGPT for Legal Services and Society},
  journal = {Michigan Technology Law Review},
  volume = {30},
  pages = {1},
  year = {2023}
}

@article{plantin2018,
  author = {Plantin, Jean-Christophe and Lagoze, Carl and Edwards, Paul N. and Sandvig, Christian},
  title = {Infrastructure Studies Meet Platform Studies in the Age of Google and Facebook},
  journal = {New Media \& Society},
  volume = {20},
  number = {1},
  pages = {293--310},
  year = {2018}
}

@techreport{pleasence2014,
  author = {Pleasence, Pascoe and Balmer, Nigel J.},
  title = {How People Resolve Legal Problems},
  institution = {Legal Services Board},
  year = {2014}
}

@article{pleasence2019,
  author = {Pleasence, Pascoe and Balmer, Nigel J.},
  title = {Justice and the Capability to Function in Society},
  journal = {Daedalus},
  volume = {148},
  number = {1},
  pages = {140--149},
  year = {2019}
}

@techreport{pleasence2015,
  author = {Pleasence, Pascoe and Balmer, Nigel J. and Denvir, Catrina},
  title = {How People Understand and Interact with the Law},
  institution = {Legal Education Foundation},
  year = {2015}
}

@article{prescott2017,
  author = {Prescott, James J.},
  title = {Improving Access to Justice in State Courts with Platform Technology},
  journal = {Vanderbilt Law Review},
  volume = {70},
  pages = {1993},
  year = {2017}
}

@article{ryan2024,
  author = {Ryan, Francine and Hardie, Liz},
  title = {ChatGPT, I Have a Legal Question? The Impact of Gen AI Tools on Law Clinics and Access to Justice},
  journal = {International Journal of Clinical Legal Education},
  volume = {31},
  pages = {166},
  year = {2024}
}

@article{sandefur2015a,
  author = {Sandefur, Rebecca L.},
  title = {Elements of Professional Expertise: Understanding Relational and Substantive Expertise through Lawyers' Impact},
  journal = {American Sociological Review},
  volume = {80},
  number = {5},
  pages = {909--933},
  year = {2015}
}

@article{sandefur2015b,
  author = {Sandefur, Rebecca L.},
  title = {What We Know and Need to Know about the Legal Needs of the Public},
  journal = {South Carolina Law Review},
  volume = {67},
  pages = {443},
  year = {2015}
}

@article{sandefur2019,
  author = {Sandefur, Rebecca L.},
  title = {Access to What?},
  journal = {Daedalus},
  volume = {148},
  number = {1},
  pages = {49--55},
  year = {2019}
}

@article{sandefur2022,
  author = {Sandefur, Rebecca L. and Denne, Emily},
  title = {Access to Justice and Legal Services Regulatory Reform},
  journal = {Annual Review of Law and Social Science},
  volume = {18},
  pages = {27--42},
  year = {2022}
}

@inproceedings{schneiders2025,
  author = {Schneiders, Eike and Seabrooke, Tina and Krook, Joshua and Hyde, Richard and Leesakul, Natalie and Clos, Jeremie and Fischer, Joel E.},
  title = {Objection Overruled! Lay People Can Distinguish Large Language Models from Lawyers, but Still Favor Advice from an LLM},
  booktitle = {Proceedings of the 2025 CHI Conference on Human Factors in Computing Systems},
  year = {2025}
}

@inproceedings{seabrooke2024,
  author = {Seabrooke, Tina and Schneiders, Eike and Dowthwaite, Liz and Krook, Joshua and Leesakul, Natalie and Clos, Jeremie and Maior, Horia and Fischer, Joel},
  title = {A Survey of Lay People's Willingness to Generate Legal Advice Using Large Language Models (LLMs)},
  booktitle = {Proceedings of the Second International Symposium on Trustworthy Autonomous Systems},
  pages = {1--5},
  year = {2024}
}

@article{shen2026,
  author = {Shen, M. Karen and Huang, Jessica and Liang, Olivia and Kim, Ig-Jae and Yoon, Dongwook},
  title = {The AI Genie Phenomenon and Three Types of AI Chatbot Addiction: Escapist Roleplays, Pseudosocial Companions, and Epistemic Rabbit Holes},
  journal = {arXiv preprint arXiv:2601.13348},
  year = {2026}
}

@article{simshaw2023,
  author = {Simshaw, Drew},
  title = {Toward National Regulation of Legal Technology: A Path Forward for Access to Justice},
  journal = {Fordham Law Review},
  volume = {92},
  number = {1},
  year = {2023}
}

@article{sommerlad2004,
  author = {Sommerlad, Hilary},
  title = {Some Reflections on the Relationship between Citizenship, Access to Justice, and the Reform of Legal Aid},
  journal = {Journal of Law and Society},
  volume = {31},
  number = {3},
  pages = {345--368},
  year = {2004}
}

@book{sommerlad2015,
  author = {Sommerlad, Hilary and Harris-Short, Sonia and Vaughan, Steven and Young, Richard},
  title = {The Futures of Legal Education and the Legal Profession},
  publisher = {Hart Publishing},
  year = {2015}
}

@article{star1999,
  author = {Star, Susan Leigh},
  title = {The Ethnography of Infrastructure},
  journal = {American Behavioral Scientist},
  volume = {43},
  number = {3},
  pages = {377--391},
  year = {1999}
}

@inproceedings{star1994,
  author = {Star, Susan Leigh and Ruhleder, Karen},
  title = {Steps Towards an Ecology of Infrastructure: Complex Problems in Design and Access for Large-Scale Collaborative Systems},
  booktitle = {Proceedings of the 1994 ACM Conference on Computer Supported Cooperative Work},
  pages = {253--264},
  year = {1994}
}

@article{strauss2021,
  author = {Strau{\ss}, Stefan},
  title = {Deep Automation Bias: How to Tackle a Wicked Problem of AI?},
  journal = {Big Data and Cognitive Computing},
  volume = {5},
  number = {2},
  pages = {18},
  year = {2021}
}

@book{susskind2023,
  author = {Susskind, Richard},
  title = {Tomorrow's Lawyers: An Introduction to Your Future},
  edition = {3rd},
  publisher = {Oxford University Press},
  year = {2023}
}

@misc{tan2025,
  author = {Tan, J. and Benyekhlef, Karim},
  title = {LegalWebAgent: Empowering Access to Justice via LLM-Based Web Agents},
  year = {2025}
}

@article{tan2023,
  author = {Tan, Jinzhe and Westermann, Hannes and Benyekhlef, Karim},
  title = {ChatGPT as an Artificial Lawyer?},
  journal = {AI4AJ@ICAIL},
  volume = {3435},
  year = {2023}
}

@techreport{trinder2014,
  author = {Trinder, Liz and Hunter, Rosemary and Hitchings, Emma and Miles, Joanna and Moorhead, Richard and Smith, Leanne and Sefton, Mark and Hinchly, Victoria and Bader, Kay and Pearce, Julia},
  title = {Litigants in Person in Private Family Law Cases},
  institution = {Ministry of Justice},
  year = {2014}
}

@techreport{unesco2025,
  author = {{UNESCO}},
  title = {Guidelines for the Use of AI Systems in Courts and Tribunals},
  institution = {UNESCO},
  year = {2025}
}

@article{wald2022,
  author = {Wald, Eli},
  title = {The Access and Justice Imperatives of the Rules of Professional Conduct},
  journal = {Georgetown Journal of Legal Ethics},
  volume = {35},
  pages = {375},
  year = {2022}
}

@incollection{weber1949,
  author = {Weber, Max},
  title = {`Objectivity' in Social Science and Social Policy},
  booktitle = {The Methodology of the Social Sciences},
  editor = {Shils, Edward A. and Finch, Henry A.},
  pages = {49--112},
  publisher = {Free Press},
  address = {New York},
  year = {1949}
}

@article{wentz2005,
  author = {Wentz, Julia},
  title = {Justice Requires Access to the Law},
  journal = {Loyola University Chicago Law Journal},
  volume = {36},
  pages = {641},
  year = {2005}
}

@article{yuce2026,
  author = {Yüce, Pelin and Dai, Xiangruo and Owens, Rebecca and Elmas, Tuğrulcan},
  title = {{ChatGPT} vs Teachers vs Students: Large-Scale Analysis of Generative {AI} Discourse in Education Communities on {Reddit}},
  journal = {arXiv preprint arXiv:2605.17712},
  year = {2026}
}

\end{document}